# Simultaneous electric and magnetic field induced nonvolatile memory

M.Quintero, A.G.Leyva and P.Levy*

Dept. Física, Centro Atómico Constituyentes, CNEA
Av.Gral Paz 1499 (1650) San Martín, Prov. de Buenos Aires, Argentina

We investigate the electric field induced resistive switching effect and magnetic field induced fraction enlargement on a polycrystalline sample of a colossal magnetoresistive compound displaying intrinsic phase coexistence. Our data show that the electric effect (presumably related to the presence of inhomogeinities) is present in a broad temperature range (300 to 20 K), being observable even in a mostly homogeneous ferromagnetic state. In the temperature range in which low magnetic field determines the phase coexistence fraction, both effects, though related to different mechanisms, are found to determine multilevel nonvolatile memory capabilities simultaneously *corresponding author: (levy@cnea.gov.ar)

Two terminal nonmagnetic nonvolatile memory capabilities and multilevel switching in oxides with perovskite or related structures received lots of attention recently. Studying thin films of $Pr_xCa_{1-x}MnO_3$, a colossal magnetoresistive manganese based oxide (manganite), Liu et al. have shown[1] that electric pulse induced resistance (EPIR) changes (previously measured on insulating $SrZrO_3$ thin films[2]) are nonvolatile, accumulative, polarity dependant and reversible. Further work revealed that several strongly correlated perovskite oxide compounds exhibit similar behavior[3] upon pulsing at room temperature (namely, above the corresponding ordering temperature), and that this switching of the resistance is related to a metallic electrode - oxide interface phenomena[4] and not to bulk properties of oxides. Though full understanding of the EPIR effect is still lacking, charge trapping near this interface is thought to be at the core of the mechanism. A simple model[5] that qualitatively accounts for experimental findings on two terminal EPIR, considers oxides as inhomogeneous media (an insulating medium that contains non percolative metallic domains). Assuming the presence of "small" metallic domains near external contact electrodes, and of a bigger one in between, key observed features can be explained.

Manganites exhibiting CMR effects have very rich phase diagrams, arising from the interplay of electronic, orbital and structural degrees of freedom. They have an intrinsic inhomogeneous paramagnetic state at high temperature, related to the presence of polarons[6]. In some cases, below an ordering temperature, still another inhomogeneous state is found[6,7], consisting of 3 types of sub micrometric regions: a paramagnetic insulating one, a highly insulating charge ordered (CO) one, and a metallic ferromagnetic (FM) one, the phase separated regime. This phase separation scenario fully develops[8] in the $Pr_yLa_{5/8-y}Ca_{3/8}MnO_3$ (PL(y)) family of compounds: the end members of which exhibit homogeneous CO ( y = 5/8 ) and FM ( y = 0 ) states, while for intermediate Pr doping there is coexistence in a broad temperature range. Small external forces can modify the relative fraction of phases in a permanent way[9] allowing memory to be encoded by, for example, applying pulses of low magnetic field (*H*) and thus producing a field induced fraction enlargement.

Tsui et al. have previously reported the occurrence of the EPIR effect in polycrystalline PL(0) and PL(5/8) at room temperature[3]. In this paper we study the PL(0.3) compound in a broad temperature range, this allows us to tackle two main questions raised by the EPIR effect in oxides: 1) In which sense inhomogeinities are essential, i.e. is the effect still observable in presence of percolative and massive metallic like regions? 2) Is it possible to imprint nonvolatile memory simultaneously by two *a priori* independent mechanisms, i.e. on surface charge trapping layers and in the volume phase fraction, driven by electric and magnetic field respectively?

Taking advantage from different competing states, we provide experimental evidence for a positive answer to both issues. A polycrystalline sample of PL(0.3) with silver paste contacts was studied. Current is injected through terminals “A” and “D” (see inset in Fig. 1), voltages are measured using a small bias current ($10^{-4}$ A) and EPIR switching is produced by applying positive (write) or negative (erase) groups of several pulses of 10 V amplitude and 2 ms duration, separated by 2 ms, also through terminals “A” and “D”. While cooling, we measure the bulk resistivity ($\rho \sim V_{BC}$) by means of 4 probes, and the interfacial voltage drop ($V_{AB}$ and $V_{BD}$) using a 3 lead configuration[3,4].

Figure 1a) shows previously reported[10] electric and magnetic properties of PL(0.3). Upon cooling from room temperature, a paramagnetic insulating regime transforms into massive CO at 210 K; some few degrees below, a phase separated state formed by FM clusters nucleated within a CO matrix produces a peak in $\rho(T)$ and a steep increase in magnetization around 200 K. This regime spans a broad temperature range, until below 80 K a mostly FM state develops.

The same piece of material is now further studied. The results for $V_{BC}(T)$ (~ ρ) shown in Fig. 1b) indicate that the sample had no significant degradation after 3 years (see peaks around 210 and 120 K). A small increase in the absolute value of ρ is observed below 200 K, related to a thermal cycling[11] effect occurring in phase separated manganites. Fig. 1b) shows $V_{BC}(T)$ measured on cooling, while applying alternatively 20 write / erase pulses every 300 seconds; small dips are related to Joule heating both at the contacts and within the sample. Figure 1c) depicts the corresponding interfacial potential $V_{BD}(T)$ obtained in the same run. Write (erase) voltage pulses determine a "high" (low) interfacial voltage drop denoted as $V_{high}$ ($V_{low}$). The thermal dependence of $V_{BD}$ resembles that of $\rho(T)$ as it is a measure of the voltage drop at the "D" interface region plus the one related to bulk properties.

The EPIR factor $(V_{high} – V_{low})/< V_{BD} >$ (where $< V_{BD} > = ( V_{high} + V_{low} )/2$ ) is a measure of the electric pulsing effect. At room temperature, we obtain an EPIR value around 5% (this change can hardly be seen in Fig. 1c) due to the scale used), close to the value reported by Tsui et al.; also in agreement with their results, no changes (except for producing extra heating) were observed in $\rho(T)$ upon pulsing. On cooling, $V_{BD}$ switches between $V_{high}$ and $V_{low}$ when pulses are applied. Careful inspection, evaluating the EPIR factor just before and after pulsing reveals that the effect is observable along the whole temperature range explored (300 - 20 K), reaching always EPIR values at least around 5%. Although they are not evaluated at equilibrium, these data provide an estimation of different regimes.

Interestingly, EPIR values remain approximately constant around 5% when cooling through inhomogeneous paramagnetic (300 – 210 K) and phase separated (200 – 100 K) regimes. An increase in the EPIR value in a temperature window around 80 - 120 K was observed, reaching around 30%, a process probably related to the percolation of FM clusters. We speculate that the effect of electric pulses may be to critically favor percolation by, for instance, enlarging FM regions. Quite remarkably, non negligible EPIR values are observed even when FM clusters already percolate throughout the sample, at the low *T* nearly homogeneous FM phase (below 80 K).

In Fig. 2 we show the effect of pulsing after cooling the sample down to 125 K with no field applied. Fig. 2a) depicts low *H* pulses applied for few minutes, and Figs. 2b) and 2c) display $V_{BC}$ ( ~ ρ) and $V_{BD}$ as a function of elapsed time, when trends of 20 write / erase voltage pulses are applied. The effect of Joule heating in $V_{BC}$ is again quite apparent, and when heat is released the initial $V_{BC}$ value is recovered. Initially, i.e. when $H = 0$, we observe a slow relaxation of $V_{BC}$ (related to the spontaneous growth of FM regions against CO[10,12]) and obtain EPIR values from around 10%.

Upon the application of $H = 0.3$ T, a decrease in ρ is observed. This jump is related to both the alignment of spins and domains, and to the enlargement[10] of the FM phase driven by *H*. When the field is turned off, ρ steeply increases, without recovering its previous value, i.e. the ρ baseline is shifted, denoting a nonvolatile effect in which memory is imprinted in the amount of FM phase. As expected, being proportional to $V_{BC,}$ the interfacial potential $V_{BD}$ shown in Fig. 2c), follows a similar trend.

Additionally, and as a key finding of this work, $V_{BD}$ tracks the EPIR effect after magnetic field pulses are turned off, mounted on the baseline determined by the FM fraction encoded nonvolatile memory. The obtained figures are slightly above 5 % after applying 0.3 T, and around 10 % after applying 0.6 T, nearly the same value as that obtained after zero field cooling.

Beyond the reason behind these specific EPIR values, the main result shown in Fig. 2 is that nonvolatile electric field induced two level switching among multilevel magnetic field determined nonvolatile memory states is feasible. At 125 K, and for the particular procedure studied (20 pulses of 10 V amplitude and 2 ms time width, separated by 2ms) we obtained a ~10% change in the electric field induced switching of the interfacial potential, while a ~25% per Tesla change was observed when magnetic field was applied. On the other hand, at room temperature and using the same procedure, EPIR is around 5%, without magnetic field dependence of the interfacial potential. Different voltage amplitudes and pulse numbers may be used to optimize additional multilevel electric field capability. The electric pulse induced effect is found to be present on a broad temperature range, being observable even in a mostly homogeneous ferromagnetic state. Different mechanisms are presumably responsible for the observed electric and magnetically induced switching, simultaneously present on the same sample. Strong electronic correlations at the oxide metal interface were recently suggested to play a major role[13] on the electric pulsing effect. Further experimental work is needed to elucidate its relation with magnetically induced memory states.

We acknowledge fruitful discussions with F.Parisi and R.Weht, critical reading of the manuscript by M.Weissmann, and support from CONICET and ANPCYT PICT Nº 03-13517. AGL is also at UNSAM, PL is member of CIC CONICET.

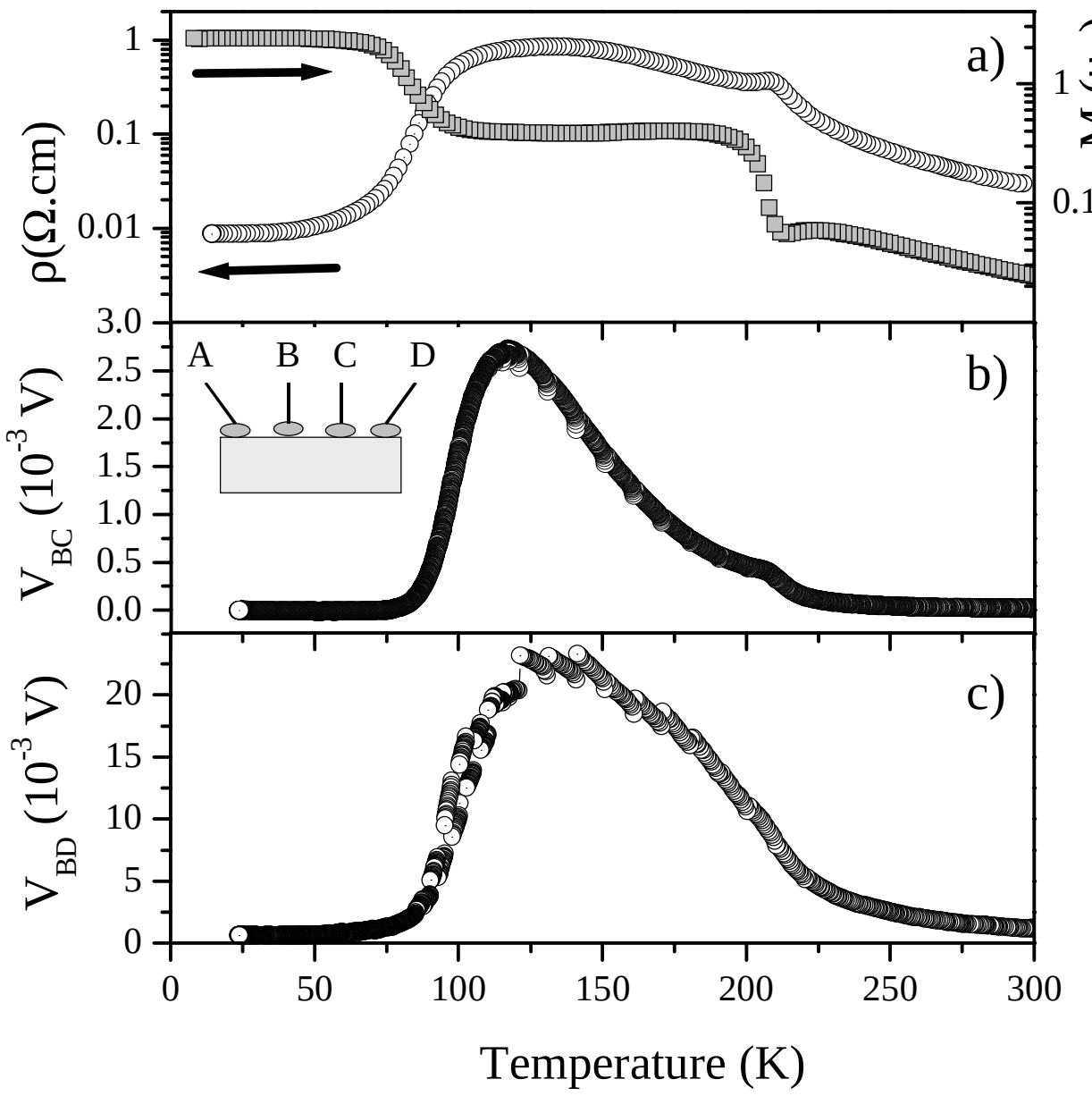

Fig.1 Temperature dependence of different properties measured on cooling: a) magnetic moment *M*(H = 0.2 T) and bulk resistivity ρ(H = 0), from Ref.10; b) $V_{BC}$ measured on the same piece of material used in a), obtained while applying alternatively 20 write or erase voltage pulses, see text; c) interfacial potential $V_{BD}$ measured in the same run. Although not always apparent due to the scale used, at all temperatures pulsing produces changes on $V_{BD}$. The inset shows the contact configuration.

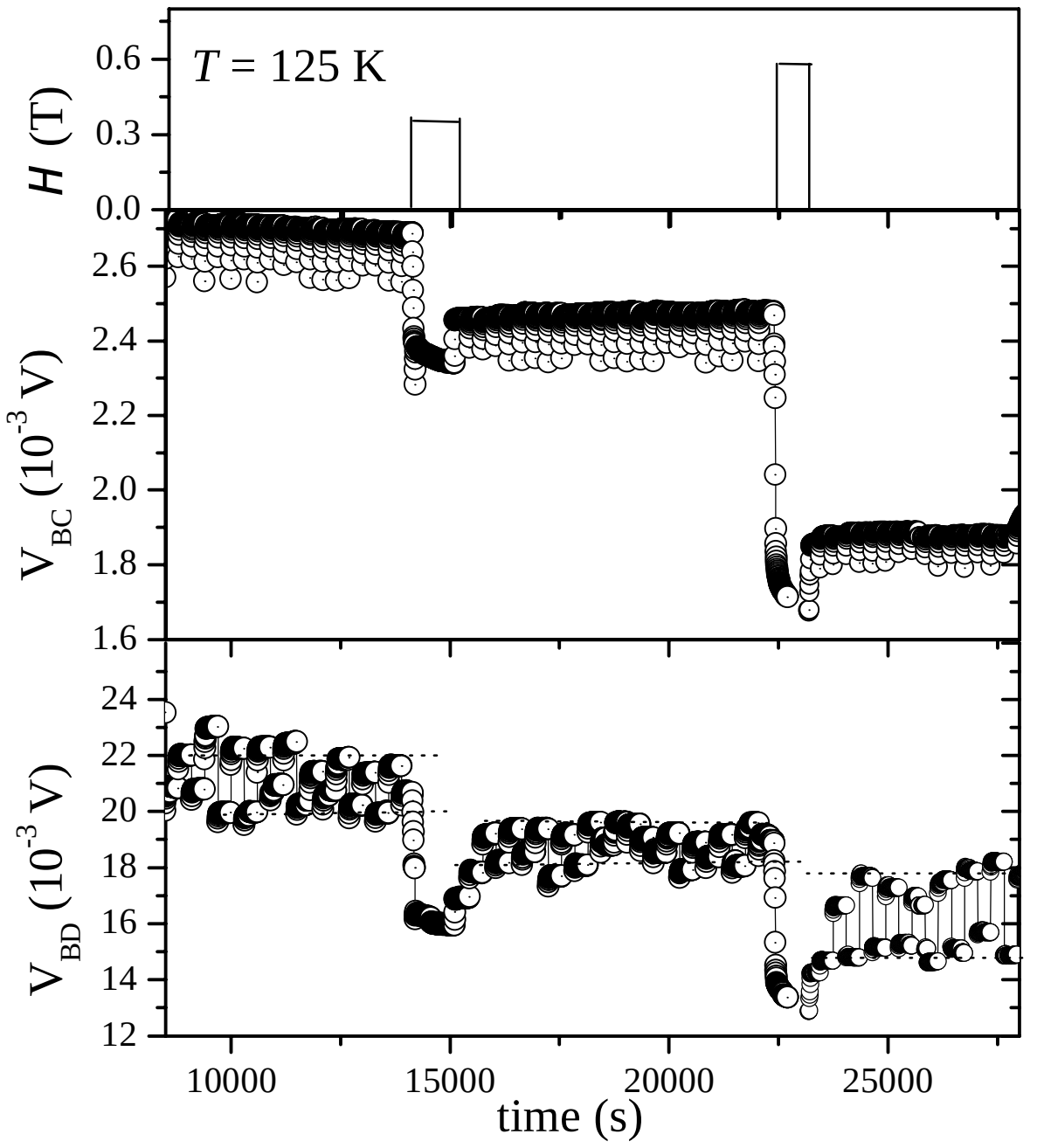

Fig.2 Time dependence of $V_{BC}$ and $V_{BD}$ after zero (magnetic and electric) field cooling to 125 K, while 20 electric field write / erase pulses are applied, in the presence of "slow" magnetic field pulses. Dashed lines depict the time averaged $V_{low}$ and $V_{high}$ values of $V_{BD}$. Electric pulses were not applied while *H* was on.